\documentclass[12pt]{article}
\usepackage{amssymb} 

\newcounter{multieqs}

\newcommand{\bq}{\begin{equation}}
\newcommand{\fq}{\end{equation}}
\newcommand{\bqr}{\begin{eqnarray}}
\newcommand{\fqr}{\end{eqnarray}}
\newcommand{\non}{\nonumber \\}
\newcommand{\noi}{\noindent}

\newcommand{\xpv}[1]{\langle #1  \rangle}

\newcommand{\rf}[1]{(\ref{#1})}

\def\alp{\alpha}       \def\gam{\gamma}
\def\del{\delta}   \def\eps{\epsilon} 
    \def\th{\theta}     
    \def\kap{\kappa}   \def\lam{\lambda} 
 \def\sig{\sigma}   
 \def\vphi{\varphi} \def\ome{\omega}

\def\Lam{\Lambda}     
\def\Ome{\Omega}

 \def\cN{{\cal N}}

\def\pa{\partial}

\def\inv{^{-1}} 

\def\pr{^{\prime}}

\def\rar{\rightarrow}

\def\one{1\!\!1\,\,}

\def\hlf{\frac{1}{2}}
\def\ove#1{\frac{1}{#1}}

\def\Box{\square} 
\def\bz{\bar{z}}
\def\ZZ{\mathbb{Z}}
\begin{document}

\thispagestyle{empty}
\setcounter{page}{0}

%%%%%%%%%%%%%%%%%%%%%%%%%%

\begin{flushright}
\begin{tabular}{l} 

CU-TP-971 \\
NIKHEF-00-006 \\
YITP-00-09  \\

\end{tabular}
\end{flushright}

\bigskip

\begin{center}

{\Large \bf Warped Compactifications in M and F Theory}\\

\vspace{12mm}

{\large Brian R. Greene${}^{*}${}\footnote{E-mail: 
greene@phys.columbia.edu},
Koenraad Schalm${}^{\sharp}${}\footnote{E-mail: kschalm@nikhef.nl} 
and Gary Shiu${}^{\dagger}${}\footnote{E-mail: shiu@insti.physics.sunysb.edu}}
\vskip 1cm
{${}^{*}$ {\em Department of Physics and Mathematics,
Columbia University \\
New York, NY 10027}} \\[5mm]
{${}^{\sharp}$ {\em NIKHEF Theory Group\\ P.O. Box 41882 \\ Amsterdam 1009DB, The Netherlands}} \\[5mm]
{${}^{\dagger}$ {\em C.N. Yang Institute for Theoretical Physics,
State University of New York\\
Stony Brook, NY 11794-3840} }

\vspace{5mm}

{\bf Abstract}
\end{center}

\noi We study M and F theory compactifications
on Calabi-Yau four-folds in the presence of non-trivial background
flux.
The geometry is warped and belongs to the class of $p$-brane metrics. 
We solve for the explicit warp factor in the orbifold limit of these 
compactifications, compare our results to
some of the more familiar
recently
studied warped scenarios, and
discuss the effects on the low-energy theory. 
As the warp factor is generated solely by backreaction, we may use 
topological arguments to determine the massless spectrum. We perform the 
computation for the case where the four-fold equals $K3 \times K3$.

\vfill

\begin{flushleft}

March 2000 

\end{flushleft}

\newpage
\setcounter{footnote}{0}

\section{Introduction}

Recently, there has been renewed 
interest
in the idea
that our world may be a domain wall in a higher dimensional 
spacetime \cite{early}. 
This scenario, often known as the ``Brane World'' 
scenario \cite{horava,add,ST,BW,ovrut,dienes},
is motivated by the existence of 
solitonic objects in string theory, such as D-branes,
in which gauge and matter fields are localized on lower-dimensional subspaces. 
Indeed, if the Standard Model fields
are localized on a brane, whereas gravity propagates in a higher dimensional
spacetime, a number of phenomenological issues have to be revisited.
A rather interesting 5-dimensional 
example was suggested in \cite{RS}
where gravity seems to be localized on a 4-dimensional 
``brane'', 
even though the 5-th dimension
has an infinite extent. 
A key ingredient is their setup is that the ambient spacetime is warped;
the localization of gravity is due to an exponentially damped warp factor.
 Several attempts have been made to find 
domain wall solutions in 5D gauged supergravity that may perhaps localize
gravity \cite{susydomain}. A warped metric can also be used to 
generate the hierarchy
between the electroweak and the four-dimensional Planck scale \cite{RS}.
The possible embedding of such a scenario
in string theory was discussed in  
\cite{hv}. 
 
In fact, warped compactifications are a rather logical generalization
of compactifications of the product type ${\cal M}_4 \times {\cal K}$, 
and were considered some time ago in
the context of supergravity \cite{peter} and heterotic string 
theory \cite{Strom,deWit}. In perturbative string theory or
supergravity there are no branes and in these scenarios there are
therefore no gauge and matter fields which are a priori localized in some
lower dimensional subspace.  
In strongly coupled heterotic string theory, on the other hand,
there are two ``end of the universe''- branes at the ends of the 
non-perturbative 11-th direction on which gauge and matter fields are
localized \cite{horava,ovrut}. In general, 
the four-form field strength $G$ is non-zero, and as a result,
spacetime is warped. 
However, the long wavelength expansion in eleventh dimensional supergravity
is valid provided that $R_{11} \leq R_{CY}/{\kappa}^{2/3}$ where 
$\kappa$ is the eleventh dimensional gravitational coupling, and
$R_{CY}$ is the size of the Calabi-Yau on which the heterotic string theory
is compactified. One cannot make the
extra dimension transverse to the branes too large, or else 
the effective field theory breaks down.

F theory \cite{vafa} provides another powerful tool in 
studying non-perturbative
string vacua (see,{\em e.g.},
\cite{F4}, for four-dimensional F theory compactifications). 
In compactifying M and F theory to lower dimensions,
anomaly cancellation (cancellation of tadpoles)
often requires one to introduce branes \cite{SVW}
(or alternatively, to turn on some background flux \cite{Becker,Sethi}). 
All $p$-brane
supergravity 
solutions are examples of warped metrics due to the backreaction of
the branes. 
The recent developments give us motivation to better understand 
these warped compactifications in a variety of circumstances.
And while the 
consistency requirements for M theory compactifications on Calabi-Yau
four-folds (and their lifting to F theory)
have been studied in some detail 
more recently \cite{Becker,Sethi,SVW,wittphase,gukov}, the
form of the background geometry
has not been determined explicitly in the presence of 
non-zero background flux (see, however \cite{verlinde2}). A purpose 
of the present paper
is to take some modest steps in this direction.

In the first part of this paper, 
we will determine the warp factor for some M/F theory
compactifications with background flux and analyze its 
behavior.
To solve the equation for 
the warp factor explicitly, we consider 
orbifold limits of Calabi-Yau spaces. 
In the F theory limit, the dilaton will also
be taken to be constant over the base (the $7$-brane charges are locally 
cancelled). 
It is, however,
likely that the generic features we find here will also hold for 
general Calabi-Yau
manifolds. We find that the background flux contributes to the warp factor in
the same manner and with 
the same scale as $p$-branes (both are $\alpha^{\prime}$
suppressed as they are due to backreaction). The resulting
metrics are thus generalizations of the $p$-brane ones 
with a contribution to the harmonic function from the background flux. 
In particular the warp factor always
contains a constant part. This is to some extent expected 
as the contribution of the background
flux to the energy momentum tensor scales inversely with the 
compactification volume. 
In the large volume limit, the contribution is
negligible, and we should recover the regular $p$-brane metric.
By way of comparison, the contribution of a 
cosmological constant (as in \cite{RS}) 
does not scale with the compactification volume.
In
the proposal of \cite{hv}, 
however, one restricts attention to the near-horizon region where
the constant part of the warp factor
is unimportant and the effective geometry changes. 
The compactification data as well as the undecoupled gravitional interactions
are encoded in a hypothetical Planck brane at the edge of the 
AdS throat.
 
In the second part of this paper we will address some aspects of the computation of (the
closed string part
of) the low-energy spectrum for compactifications with background flux. 
Generic
string compactifications suffer from an abundance of massless
fields. The presence of background flux, however, will lift a number of these
\cite{Sethi}. {\em A priori}, 
the number of massless fields are determined by the moduli of the
Calabi-Yau four-fold, subject to a set of topological
consistency conditions on the non-zero background flux. 
On the other hand, a Kaluza-Klein reduction of an arbitrary warped metric
 generates
 extra mass terms due to the warp factors. Generically 
the massless spectrum is no longer determined purely from the dimensions of 
spaces of harmonic forms on the internal space.
However, 
the powers of the
warp factors that appear in this type of warped compactifications, 
which are determined by supersymmetry, balance each other
out, so one may indeed use topological information to determine
the massless spectrum.
Moreover the non-trivial parts of the warp
factor are generated solely by the gravitational backreaction, and in the
weak coupling limit space-time reduces back to a product form. 
Using this information we recover explicitly
the superpotential for the complex structure moduli conjectured 
in \cite{gukov}.

The organization of the paper is as follows.
In Section \ref{review}, we review the results of 
M and F theory compactifications on Calabi-Yau 4-folds 
in the presence of background fluxes.
In Section \ref{warp}, we determine the warp factor for
orbifold examples, in which the anomaly is cancelled by a combination
of background flux (from the untwisted and twisted sectors) and 
$2/3$-branes. 
In Section \ref{shape}, we comment on the shape of the graviton wavefunction
in this type of warped compactification. For completeness,
we include in the appendix
a parallel discussion on the graviton wavefunction
for Heterotic-M theory on a Calabi-Yau 3-fold.
In Section \ref{moduli}, we present a general discussion
on determining the number of complex and Kahler moduli in vacua
with background flux.
For illustative purpose, we consider in Section \ref{tops} compactifications
of M theory with background
flux which satisfy the field equations
for an orbifold limit of $K3 \times K3$. The conditions on the
background flux are greatly
simplified if we choose it to be a product of $(1,1)$ forms of each
$K3$. For this example we will determine the spectrum using 
the explicit topological data and we will discuss how to solve more
general cases.
We end with some comments in Section \ref{discussion}.

\section{M and F theory Vacua with Background Fluxes}\label{review}

Warped compactifications of perturbative 
string theories, or rather supergravities, 
 preserving minimal supersymmetry in four
dimensions were considered in \cite{Strom,deWit}. The supersymmetry
requirements are highly restrictive; it was found that for Type II
theories there are no non-trivial solutions if the four non-compact
dimensions are Minkowski \cite{deWit}. For type I/heterotic theory
there is a solution if the internal space has nonzero torsion:
{\em i.e.} the vacuum expectation value of the three-form NS-NS 
field strength
$\xpv{H^{NS}}$ is non-zero. In this case the warp factor equals the dilaton,
whose profile is determined by the vacuum expectation value $\xpv{H^{NS}}$.

It has since been discovered that the M theory effective action has at
the first   
subleading order in the derivative expansion a topological term
\cite{X8term}
\bq
-\int C \wedge X_8(R) ~,
\fq 
where 
\begin{equation}
X_8 = {1\over 8\cdot 4!} \left( \mbox{tr} R^4 - {1\over 4} (\mbox{tr} R^2)^2 \right) ~.
\end{equation}
The inclusion of such
higher derivative terms in the
low energy supergravities allows for new solutions
preserving minimal supersymmetry, i.e. four supercharges, which are of
the warped 
kind \cite{Becker,Sethi}.
(see also \cite{SVW,wittphase,gukov}). Let us give a brief review.

On compactifications down to three dimensions, the eight-form $X_8(R)$ can take
on a background expectation value. This will act as a source term for the
three-form field $C$. To maintain a solution to the field
equations one needs to introduce M2-brane sources or 
non-trivial $G$-flux; $G$ is the
four-form which locally equals $dC$. Either the M2-branes or the
non-trivial $G$-flux or both 
induce in turn a warping of the metric \cite{Becker}  
\bq
ds^2 = e^{-\phi(y)} \eta_{\mu\nu}dx^{\mu}dx^{\nu} + e^{\hlf \phi(y)}
g_{a\bar{b}}dy^ad\bar{y}^{\bar{b}} ~.
\label{M&M}
\fq
The requirement that minimal supersymmetry (four
supercharges, $\cN=2$ in $d=3$) is preserved determines the relative
weights of the warp-factors, as is known from $p$-brane solutions
\cite{Duff}. In addition, supersymmetry demands that
$g_{a\bar{b}}$ is the metric of a Calabi-Yau 
four-fold. The four-form $G$ must furthermore obey
\bq
G_{abcd}=0=G_{abc\bar{d}}~, ~ g^{c\bar{d}}G_{a\bar{b}c\bar{d}} = 0 ~,~
G_{\mu\nu\rho a} = \eps_{\mu\nu\rho}\pa_a e^{-\frac{3}{2}\phi} ~.
\label{cond}
\fq
The first two conditions mean that $G$ is a (2,2) form on $CY_4$ which is
self-dual or, equivalently, primitive with respect to the Kahler form 
$J$ of the $CY_4$ \cite{Sethi}
\bq
G = \ast G  ~~\Leftrightarrow~~ J \wedge G =0 ~.
\fq
Dirac quantization requires $G$ to be an element of integer cohomology.
The last equation of \rf{cond} says that the only nonvanishing
part of the three-form $C$ is 
\bq
C_{\mu\nu\rho}= \eps_{\mu\nu\rho} e^{-\frac{3}{2}\phi}~.
\fq
By definition the antisymmetric tensor is taken with respect to the unwarped
metric.\footnote{The three-form thus equals the induced volume on the
three-dimensional space 
\bqr \nonumber
C_{\mu\nu\rho} &=& \sqrt{\det
|g_{MN}\pa_{\mu}z^M\pa_{\nu}z^N|}\eps^{\footnotesize
Levi-Civita}_{\mu\nu\rho}
\fqr
as is again familiar from $p$-branes. For the Dirac-delta
function in curved space, we use the convention that
$\int dx \sqrt{g(x)} \delta (x-x_i)=1$.} 

The field equation for $C$ determines the warp factor \cite{Becker,Duff}
\begin{equation}\label{Mwarp}
\Box_{CY_4} (e^{3 \phi/2}) = \ast_{CY_4} 
\left( X_8 - {1\over 2} G \wedge G \right)- \sum_{j=1}^{n}
\delta^8 (y-y_j) ~.
\end{equation}
The factor $G\wedge G$ has its origins in the well known $C\wedge G\wedge G$
supergravity interaction and we have introduced $n$ M2-branes at
arbitrary points. This number $n$ is determined by the consistency
requirement - the absence of tadpoles or 
anomalies - that the integral over the right hand side of
(\ref{Mwarp}) vanish,
\bq
\int_{CY_4} \!\!\!X_8 = \frac{\chi_{CY_4}}{24}= n+ \hlf\int_{CY_4} \!\!\!G \wedge
G ~.
\fq

\subsubsection*{\it Lift to F theory}

Such compactifications can be lifted to F theory if the $CY_4$ is
elliptically fibered. 
 Four-dimensional Lorentz-invariance requires
that the four-form $G$ has one leg on the toroidal fiber and three on
the base $\cal{B}$ \cite{Sethi}. 
We will limit our attention to the case where
the $G$-flux is not localized in the fiber of the
elliptic fibration, but is constant. In that case the
M theory solution can straightforwardly 
be lifted to F theory with \cite{Sethi}
\bqr\label{Fmetric}
ds^2 &=& e^{-\frac{3}{4}\phi}dx^2 +
e^{\frac{3}{4}\phi}g^{\cal{B}}_{a\bar{b}}dy^ad\bar{y}^{\bar{b}} ~,\non
H^{NS} &=& \ome -\ast_{\cal{B}}{\ome} ~,\non
H^{RR} &=& \ome\tau-\ast_{\cal{B}}{\ome}\bar{\tau} ~,\non
D^{+}_{\lam\mu\nu\rho} &=&\eps_{\lam\mu\nu\rho}e^{-\frac{3}{2}\phi}~,
\fqr
where $\ome\, \eps \, H^{(1,2)}(\cal{B})$ with
$(\tau-\bar{\tau})\int_{\cal{B}} \ast \ome \wedge \ome = \int_{CY_4}
G\wedge G ~\eps~ \ZZ^+$. The metric and four-form potential are again
that of the D3-brane solution. 
Formally the warp factor is still given by (\ref{Mwarp}), but one has
to be careful with the dependence on the internal directions. 
We will solve for the warp factor for
a four-fold of the form $K3 \times T^4/\ZZ_2$ where the fiber belongs
to $T^4/\ZZ_2$. In this case,
the warp factor can equivalently be 
determined from the IIB field equation for
$D^{+}$, 
\bq\label{warpeq}
d \ast d D^{+} = \ove{16}\sum_{i=1}^4 \mbox{tr}(R \wedge R)
\del^2(z^1 -z^1_i) -H^{NS} \wedge H^{RR}   - \ast_{\cal{B}} \sum_{j=1}^n \del^2(z^1 -z^1_j)\del^4(\ome
-\ome_j) ~.
\label{eqq}
\fq
Here, $i$ labels the fixed points of $T^2/\ZZ_2$ and $j$ labels
the $n$ D3-branes. Tadpole cancellation requires that this number equals
\bq
\int_{\cal{B}}\!\! H^{NS} \!\wedge \!H^{RR} +n = \frac{\chi_{K3 \times
K3}}{24} = 24 ~.
\fq
Substituting the background expectation value for $D^+$ we
find the F theory analogue of \rf{Mwarp}
\bq
\Box_{\cal{B}}(e^{3\phi/2}) = \ast_{\cal{B}}\left(\ove{16}\sum_{i=1}^4 \mbox{tr}(R \wedge R)
\del^2(z^1 -z^1_i)- H^{NS} \wedge H^{RR} \right)
- \sum_{j=1}^n \del^2(z^1 -z^1_j)\del^4(\ome
-\ome_j) ~,
\label{eqq2}
\fq
which determines the shape of the warp factor. 
In the orientifold limit, 
the first term of the expression contains the contribution of the orientifold
planes \cite{Das}.

\bigskip

We will solve the warp-factor equations \rf{Mwarp} and \rf{eqq2} for
an orbifold limit of the Calabi-Yau. Specifically we will concentrate
on the case $K3 \times K3$ $= T^4/\ZZ_2 \times T^4/\ZZ_2$. 
This allows us to take
the F theory limit in the end, for which we will then attempt to
compute the massless four-dimensional spectrum. 

Let us note from the outset that the warp factor, as a solution to 
(\ref{Mwarp}),(\ref{eqq2}) is determined only up
to an integration constant whose value is
fixed by the boundary conditions of the warp factor. If the fluxes are
sufficiently localized, the metric
\rf{M&M} should be approximately flat away from such special points 
and the constant
can be fixed to $1$. 
We will discuss the effect of a constant flux in the next section. 

\subsubsection*{\it The Dilaton}

In the previously known warped compactifications of (heterotic) string theory,
the warp factor turned out to be equivalent to the dilaton
\cite{Strom,deWit}. 
Let us
briefly recall why this is
not the case for the Type IIB dilaton when we
lift the above M theory vacua to 
F theory \cite{Sethi}. 
Weyl rescaling between the Einstein and the string frame can
therefore alter the form of the warp factors if a non-constant dilaton
profile is consistent with the field equations. 

By compactifying M theory in the eleventh and ninth-direction to
 IIA on $S_1$ and T-dualizing on the latter one finds that the 
IIB dilaton is given by a ratio of the compactification radii \cite{Aspinwall}
\bq
\exp{(\vphi_{IIB})} =\frac{R_{11}}{R_9}~.
\fq
Since in the background solution for an ellipitically fibered $CY_4$,
\bq
ds^s = e^{-\phi} dx^2 + e^{\phi/2}(g_{a\bar{b}}dy^ad\bar{y}^{\bar{b}} 
+ R_9^2(y)dw_1^2+R_{11}^2(y)dw_2^2) ~,
\fq
the warp factor is common to both $R_{11}$ and $R_9$, the IIB dilaton is 
independent hereof. In general, though,
 the dilaton will be a function of the internal dimensions $y_a$. 
In F theory, the $7$-branes are sources
of the complex field $\tau=a+i e^{-\varphi}$ where $a$ is the RR axion.
For example, near the location of a $7$-brane at $z=z_i$ ($z$ is 
the complex coordinate transverse to the $7$-branes),
\begin{equation}\label{monodromy}
\tau \sim {1\over {2 \pi i}} \mbox{log} (z-z_i) ~.
\end{equation}
In the Einstein frame the metric is then warped to
\cite{greene,vafa}
\begin{equation}\label{stringycosmic}
g_{z\overline{z}} = \tau_2 \eta^2 \overline{\eta}^2 \prod_{i} (z-z_i)^{-1/12}
\prod_{i} (\overline{z}-\overline{z}_i)^{-1/12} ~.
\end{equation}
In F theory, the D7-branes are not mutually local, but  
in the orientifold limit, which we will consider, the charges of the D-branes
are locally cancelled against the orientifold planes (which for non-zero $g_s$, are bound
states of some $(p,q)$ $7$-branes). In that case the complex
field $\tau$ becomes a constant, and the metric (\ref{stringycosmic}) reduces
to a flat one (except for conical 
singularities at the locations of the orientifold
planes). In the general case where the $7$-brane
charges are not locally cancelled, the calculation of the
effective four dimensional
Planck scale would involve an integral over a rather non-trivial 
function of the internal manifold 
(due to (\ref{monodromy}), (\ref{stringycosmic})).
Another complication is that the metric for the 3-7 brane system
where the 7-brane charges are not locally
cancelled has not yet been solved; for progress 
in this direction see \cite{37system}.

\section{The Warp Factor}\label{warp}

\subsubsection*{\it M theory}

The warp factor is to be determined from the equation of motion 
(\ref{Mwarp}) for
the gauge field $C_{\mu\nu\rho}$. To find its explicit form, 
we have to invert the Laplacian on the
compact internal space. This can be done with the help of Green's
functions. On a compact space, or equivalently when the Laplacian has
non-trivial zero modes, the inversion is only defined on the subset of
functions orthogonal to these zero modes, i.e. on those functions
not belonging to
the kernel of the Laplacian (see appendix for a brief review). 
For our purposes, it is relevant to know that on a
compact space the scalar Green's function obeys,
\bq
\Box G(x,x_i) = \del^d(x-x_i)-\frac{1}{\mbox{Vol}}
\fq
where $\del^d(x-x_i)$ is the $d$-dimensional Dirac delta-function, 
and ``Vol'' is the volume of the compact space.

On an orbifold, internal fluxes fall into two categories: those
localized at the fixed points and those wrapped
over periods of the underlying torus.   
The equation \rf{Mwarp} determining the warp factor thus has
two different kinds of contributions: the part of the flux term
$G \wedge G$
 which consists of a {\em constant}
flux corresponding to ``untwisted'' cycles and that part which is built
from 
``twisted'' cycles, proportional to Dirac-delta-functions at the fixed
points. 
The constant fluxes are naturally
expressed as an integer divided by the volume (after the Hodge dual
has been taken). In our case, the orbifold limit of $K3 \times K3$, 
we take the G-flux
to be constant on one of the K3's, so that we may take the F theory
limit later.   Then the flux term on the r.h.s. of \rf{Mwarp} 
can be written as
\bq
\ast \left(G\wedge G\right) = \frac{r_1r_2}{V_1V_2} +\frac{r_1}{V_1} \sum_{x_p}
m_p\del^4(x-x_p) ~,
\label{mmm}
\fq
where $r_i$, $V_i$ is respectively
the total untwisted flux and volume on each $K3$ 
and $m_p$ is the total twisted flux
located at each fixed point $x_p$ of the second $K3$. 

In the orbifold limit the curvature is also localized at the fixed
points. On $K3 \times K3$ the total contribution to the warp factor from
the curvature is
\bq
\int X_8 =\frac{\chi_{K3\times K3}}{24} = 24 ~.
\fq
The $(T^4/\ZZ_2)^2$ limit has $16^2$ fixed points. Each 
contributes of course 
equally to the curvature and the first term in \rf{Mwarp}
may thus be written as
\bq
\ast X_8 = \frac{24}{16^2}\sum_{x_p,z_p} \del^4(x-x_p)\del^4(z-z_p) ~,
\fq
where $z_p$ are the fixed points of the first $K3$.

Combining this information the equation \rf{Mwarp} determining the
warp factor reduces to
\bqr
\Box(e^{3\phi/2}) &=& \frac{3}{32}\sum_{x_p,z_p} \del^4(x-x_p)\del^4(z-z_p)
-\frac{r_1r_2}{2V_1V_2} 
-\frac{r_1}{2V_1} \sum_{x_p}m_p\del^4(x-x_p) \non
&&
- \sum_{i=1}^n \del^4(x-x_i)\del^4(z-z_i) \non
&=& \frac{3}{32}\sum_{x_p,z_p}
\left(\del^4(x-x_p)\del^4(z-z_p)-\ove{V_1V_2}\right) 
-\sum_{i=1}^n\left(\del^4(x-x_i)\del^4(z-z_i)-\ove{V_1V_2}\right) \non
&&
-\frac{r_1}{2V_1} \sum_{x_p}m_p\left(\del^4(x-x_p)-\ove{V_2}\right) ~.
\fqr
Here we have made use of the fact that tadpole cancellation implies
that $48=r_1r_2+r_1\sum m_p+2n$. 

The warp factor is now easily expressed in terms of Green's functions
\bqr
e^{3\phi/2} = c_0 +
\frac{3}{32}\sum_{x_p,z_p}G^{(8)}(x,(x_p,z_p))-\sum_{x_i,z_i}G^{(8)}(x,(x_i,z_i))-\frac{r_1}{2V_1}\sum_{x_p}
m_pG^{(4)}(x,x_p) ~.
\label{warpie}
\fqr
Note that the completely constant flux term, proportional to $r_2$,
does not contribute to the warp factor in any obvious way. 
The Green's functions $G^{(8)}$ on $K3\times K3$ and $G^{(4)}$ on
a single $K3$ can be constructed in the orbifold limit by the method
of images. For example, $G^{(4)}$ is given by 
\begin{equation}
G^{(4)} (x,x_i) = - \sum_{\vec{p}} \frac{e^{i \vec{p} (\vec{x}-\vec{x_i})}
+e^{i \vec{p} (\vec{x}+\vec{x_i})}}{\vec{p}^2} ~.
\end{equation}
The momenta $\vec{p}$ are quantized in units of the inverse radii
of the $T^4$. The second term is due to the $\ZZ_2$ image.
In the above sum, the zero momentum mode ($p_1=p_2=p_3=p_4=0$) is
excluded.

The integration
constant $c_0$ is determined by the boundary conditions. We
already explained why, if all fluxes are localized ($r_2=0$ in our
example above) this constant is fixed to unity. In that case far away
from the points where flux (energy density) is localized, spacetime
should be approximately flat. 
In the case of constant G-flux over
the whole internal manifold, we can determine $c_0$ from the
curvature in the internal directions by looking at the Einstein
equation,
\bq
R_{MN} = -\hlf\left(G_{M}\cdot G_{N} - \frac{3}{D-2} g_{MN}G\cdot G\right) ~,
\fq
where
\bq
G \cdot G = \frac{G_{MNRS}G^{MNRS}}{4!}~~~~~~;~~~~~~ G_{M} \cdot G_{N}
= \frac{G_{MABC}G^{~~ABC}_N}{3!} ~.
\fq 
We have ignored contributions to the stress-tensor from the curvature
and the M2-branes as they are localized. Substituting the background
expectation values (and noting that $G$ is self-dual in
the internal dimensions)
\bq
C_{\mu\nu\rho} =\eps_{\mu\nu\rho}e^{-3\phi/2} ~~~~,
~~~~~~~G_{a\bar{b}c\bar{d}} \equiv \tilde{G}_{a\bar{b}c\bar{d}} \neq 0
~~\rar ~~\tilde{G} \cdot\tilde{G} = r/V_{CY} ~,
\fq
one finds for the Ricci curvature in the internal
directions\footnote{To find this answer one needs to use that
the field equations require that
\bqr \nonumber
\tilde{G}_m \cdot \tilde{G}_n = \frac{g_{mn}}{2}
\tilde{G}\cdot\tilde{G} ~.
\fqr
This is the combined constraint of primitivity and $G ~\eps~ H^{(2,2)}$
expressed in components.},
\bq
R_{mn} = \hlf\left({\pa_m\ln e^{3\phi/2}}{\pa_n\ln e^{3\phi/2}}
-\frac{g_{mn}}{3}(\pa \ln e^{3\phi/2} )^2 + g_{mn}\frac{e^{-3\phi/2}}{6}
\tilde{G}\cdot \tilde{G} \right) ~,
\fq
with $g_{mn}$ the metric on the $CY_4$. In the decompactification
limit the explicit $G$-flux term vanishes as it scales inversely with
the volume. The warp factor does not depend explicitly 
on the constant flux and
we should therefore recover the regular $p$-brane metric for which
$c_0$=1. (Strictly speaking we have only shown that $\lim_{V_{CY} \rar
\infty} c_0(V) =1$). 
The above argument is rather
general, and is not restricted to orbifold cases where the curvatures
are localized. This is again because $\int X_8$ and $\int G \wedge G$
are quantized (to
a finite number), and so their contribution to the internal curvature is
small when the size of the Calabi-Yau is large. In other words, both
$\int X_8$ and $\int G \wedge G$, unlike the cosmological constant,
are not extensive quantities.
By way of comparison, a non-zero $c_0$ means that the decompactification
limit of the present scenario has significantly different properties than those
studied in \cite{RS}. For instance, one would not obtain a finite
Planck scale in this limit.
   
An issue regarding the solution \rf{warpie}
is that due to the opposite sign
of the Green's functions corresponding to the curvature induced
charge, the metric has a naked singularity at the fixed points of 
the orbifold: the conical orbifold singularity. This is,
however, an artifact of the long-range supergravity
approximation. The solution 
may roughly only be trusted as long as the distance to any special point
is larger than the Planck length or string scale. The naked singularity
is expected to be cured by stringy effects. 
Examples of this have recently been
discussed in \cite{polch,john}. Since these effects will only modify
the metric close to the singularity, they are unlikely to change the
arguments above.

\subsubsection*{\it F theory}

For F theory the solution is similar.
We denote $V_6$ as the volume of the 6 compactified dimensions and 
$V_2$ the volume of the two dimensions
transverse to the $7$-branes. The twisted fluxes are proportional
to $\delta^4(x-x_p)$ since we consider only fluxes that are not
localized around a singular fiber of the elliptic fibration: constant
on one of the K3's.
The flux factor can thus be written as
\bq
\ast(H^{NS} \wedge H^{RR} ) =  \frac{r}{V_6} 
+ \frac{1}{V_2} \sum_{x_p} m_p\del^4(x-x_p) ~.
\label{nnn}
\fq
 
The curvature contribution to the equation for the warp factor is due
to the (unit charge) D3-branes or O3-planes (charge $\mu_O$). In the
orbifold limit the curvature is completely localized at the fixed
points and thus gives rise to a six-dimensional delta-function. This
corresponds to the fact that we are considering 
limit that the D7-brane
charges 
are locally cancelled and we are left with only O3-plane charge at the
fixed points. Explicitly
\bqr
\Box e^{\frac{3}{2}\phi} &=& \sum_{x_O}\mu_O\del^6 (x-x_O)-
\ast(H^{NS}\wedge H^{RR}) 
- \sum_{i=1}^n \del^6 (x-x_i) ~.
\fqr
Substituting \rf{nnn} we can rewrite this as 
\bqr
\Box e^{\frac{3}{2}\phi} &=& \sum_{x_O}\mu_O\del^4 (x-x_O) \del^2
(z-z_O)
- \frac{r}{V_6} 
- \frac{1}{V_2} \sum_{x_p}m_p\del^4(x-x_p)
                    - \sum_{i=1}^n \del^6 (x-x_i) \nonumber \\
&=& \sum_{x_O}\mu_O \left( \del^4 (x-x_O) \del^2 (z-z_O) - \frac{1}{V_6} \right)
-\frac{1}{V_2} \sum_{x_p} m_p \left( \del^4(x-x_p) - \frac{1}{V_4} \right) 
\nonumber \\
&& - \sum_{i=1}^n \left( \del^6 (x-x_i) -\frac{1}{V_6} \right) ~,
\fqr
where we have made use of the fact that anomaly cancellation implies
$r=n+ \sum_{x_p} (m_p + \mu_{O})$.
In terms of Green's functions the warp factor is:
\bq
e^{\frac{3}{2}\phi} = c_0 - \frac{1}{V_2} \sum_{x_p} m_p G^4 (x,x_p)
+ \sum_{x_O} \mu_{O} G^6 (x,x_O) - \sum_{i=1}^{n} G^6(x,x_i) ~.
\label{rrr}
\fq

For the more general case 
where the $7$-brane charges are
not locally cancelled, 
equation (\ref{warpeq}) (which is written in the
string frame), should receive contributions from O3- and
O7-planes and D3- and 
D7-branes plus an appropriate contribution from the 
dilaton
which is no longer constant. 
The net effect is that, in the Einstein frame,
the warp factor is given by an equation of the form
\begin{equation}\label{warp37} 
\partial_i \left( \sqrt{g} g^{ij} \partial_j e^{3 \phi/2} \right)
= \sum_{x_O}\mu_{O}\del^6 (x-x_O) 
-H^{NS} \wedge H^{RR}
- \sum_{i=1}^n \del^6 (x-x_i) ~,
\end{equation}
where the metric $g_{ij}$ takes into account the backreaction of
the $7$-branes according to (\ref{stringycosmic}). 
The exact solution
to this equation is not known, although it can be solved
approximately \cite{37system}. 

\section{The Shape of Gravity}\label{shape}

With the warped metric (\ref{Fmetric}) in F theory, 
the tree-level four-dimensional Planck scale is given in terms of 
the 10-dimensional Planck scale by
\begin{equation}\label{M4}
M_{4}^2 = M_{10}^8 \int_{\cal{B}} e^{-2 \varphi} e^{3 \phi/2} 
\sqrt{g^{\cal{B}}} ~,
\end{equation}
whereas the tree-level gauge coupling on a three-brane
at $y=y_i$ has no warp factor contribution and is given as usual by
\begin{equation}
{1\over g_{YM}^2(y_i)} = e^{-2 \varphi (y_i)} ~.
\end{equation} 
As the non-trivial terms in the warp factor are solely due to backreaction, 
the usual relation between the four-dimensional
Newton constant and the three-brane
gauge coupling still holds at leading order.

At subleading order, the warp factor might come into play.
However, an explicit calculation shows that this is not the case.
First, let us consider the case where the dilaton is constant.
Note that the power of the warp factor that appears in
(\ref{M4}) is exactly equal to the one in the
field equation (\ref{eqq2}). Since the Green's function may be
written as 
\begin{equation}
G(y,y_i) = - \sum_{\lambda \neq 0} \frac{\bar{f}_{\lambda} (y) 
f_{\lambda} (y_i)}{\lambda^2} ~,
\end{equation}
where $f_{\lambda}$ are orthonormal eigenfunctions of the Laplacian with
eigenvalues $\lambda$, 
the four-dimensional
Planck mass can be expressed as
\begin{equation}\label{M2correct}
M_4^2 = M_{10}^8 e^{-2\varphi} \left( \int d^6 y \sqrt{g^{\cal{B}}} - 
\sum_{n \neq 0} {1\over {\lambda^2}}  \int d^6 y^{\prime} \sqrt{g^{\cal{B}}}
\rho (y^{\prime}) 
f_{\lambda} (y^{\prime})
\int d^6 y 
\sqrt{g^{\cal{B}}} \bar{f}_{\lambda} (y) \right)
~.
\end{equation}
The quantity $\rho$ equals the charge density on the r.h.s. of (\ref{eqq2}),
\bq
\rho (y^{\prime}) = 
\ast_{\cal{B}}\left(\ove{16}\sum_{i=1}^4 \mbox{tr}(R \wedge R) 
\del^2(y^{\prime} -y_i)- H^{NS} \wedge H^{RR}  \right)
- \sum_{j=1}^n \del^6(y^{\prime} -y_j) ~.
\fq
Because the zero mode is explicitly excluded in the Green's function,
the second term in (\ref{M2correct}) vanishes. 
This bears out our expectations. 
The warp factor does not change
the relation between the four and ten-dimensional Planck scale.\footnote{
If the dilaton is not constant (taking into account the $7$-branes), 
the net effect is the
modification of the field equation to (\ref{warp37})
where the metric 
is expressed in the Einstein frame. The above argument
based on the properties of the
Green's function should also hold in this case.}

In \cite{RS}, an exponentially decaying wavefunction of the graviton
was used to generate the hierarchy between the electroweak scale and the 
four-dimensional 
Planck scale. The exponential decay is due to the fact that the
wavefunction of the massless graviton
is dressed by some powers of
the warp factor. By way of comparison,
let us also deduce the shape of the
graviton wavefunction in the present setup by
linearizing fluctuations about the background warped metric
in Section \ref{warp}. We will see that although the massless spectrum
is indeed unaffected, the masses of the KK modes
will get dressed by powers of the warp factor.
A few comments are in order. 
If the compactification is of the order of string scale, we expect,
{\em a priori}, that the wavefunctions of the KK modes
are no longer given by linearizing Einstein gravity since stringy
effects may become important.
However, in compactifying Heterotic-M theory
on a Calabi-Yau three-fold \cite{ovrut}
(the G flux in this case is generically non-zero), there is a regime
in which the theory is effectively five-dimensional, the wavefunctions
of some of the lower-lying KK modes (in the fifth dimension)
can still be obtained by linearizing Einstein gravity.
The analysis for 
Heterotic-M theory on $CY_3$ is very similar,
so for completeness, we have included a derivation of the KK spectrum 
in the appendix.

For a metric of the form
\begin{equation}
ds^2 = e^{2A(y)} \eta_{\mu\nu} 
dx^{\mu} dx^{\nu} + e^{2B(y)} \overline{g}_{ab} dy^a dy^b ~,
\end{equation}
the gravitational fluctuations are given by
$\eta_{\mu \nu} \rightarrow \eta_{\mu\nu} + h_{\mu\nu}$
where $h_{\mu \nu}$ is small compared with
$\eta_{\mu \nu}$.

From Einstein's equations, if we choose the gauge 
$\partial^{\mu} h_{\mu \nu}=0$, 
the
linear fluctutations can be shown to
satisfy the
covariant wave equation (see the discussion in the next section or
 {\em e.g.} \cite{csaki}):
\begin{equation}\label{covariantwave}
{1\over \sqrt{g}} \partial_{M} \left( \sqrt{g} g^{MN} \partial_{N} h_{\mu\nu} \right) = 0 ~,
\end{equation}
where the indices $M,N=0,1,\dots,9$ 
are raised and lowered with the warped background metric (in Einstein frame). 
For the 
warped metric in M and F theory that we considered ({\em i.e.},
$A(y)$ is proportional to $B(y)$), this reduces to
\begin{equation}
\left( e^{-2 A(y)} \Box_{Mink} + \Box_{\bar{g}} \right)h_{\mu\nu}  = 0 ~.
\end{equation} 
Expanding $h_{\mu\nu}(x,y) = \psi (y) \hat{h}_{\mu\nu} (x)$,
with $\Box_x \hat{h}_{\mu\nu} (x) = m^2 \hat{h}_{\mu \nu}(x)$, we have
\begin{equation}\label{wave}
\Box_{\overline{g}} ~\psi (y)
= - m^2 e^{-2A(y)} \psi (y) ~.
\end{equation}
Hence the masses of the KK modes depend on the warp factor, which
in turn depends on the locations of the branes.

For the 
massless graviton ({\em i.e.}, $m^2=0$), (\ref{wave}) always admit the
solution $\psi(y)=$ constant. The wavefunction
for the graviton, however, should be properly normalized:
\begin{equation}
S \sim \int d^{10} x \sqrt{g} \partial_{\lambda} h_{\mu \nu} 
\partial^{\lambda} h_{\mu \nu} +\dots =
\int dy ~e^{-2A(y)}  \sqrt{g(y)} \psi^2(y) \cdot
\int d^4 x \partial_{\hat{\lambda}} \hat{h}_{\mu\nu} (x) 
\partial^{\hat{\lambda}} \hat{h}_{\mu\nu} (x) + \dots
\end{equation}
where hatted indices are with respect to the unwarped metric.
Therefore, the properly normalized wavefunction is
\begin{equation}
\Psi (y) = \left[ e^{-2 A(y)} \sqrt{g(y)} \right]^{1/2} \psi(y)
= e^{- \varphi} e^{3 \phi/4} ~ \left[ det~g_{ab}^{\cal{B}} \right]^{1/4} ~,
\end{equation}
which is the square root of the integrand in (\ref{M4}), as expected.
For the orbifold examples that we discuss in Section \ref{warp}, 
the dilaton
$\varphi$ and 
$\left[ det~g_{ab}^{\cal{B}} \right]^{1/4}$
 are constant, 
with $e^{3 \phi /2}$ 
given in terms of the Green's functions, hence
\begin{equation}\label{graviton}
\Psi (y) \sim \left[ c_0 - \frac{1}{V_2} \sum_{x_p} m_p G^4 (x,x_p)
+ \sum_{x_O} \mu_{O} G^6 (x,x_O) - \sum_{i=1}^{n} G^6(x,x_i) \right]^{1/2}~.
\end{equation}

An alternative view of these warped compactifications
was suggested in \cite{hv}.
In extending the AdS/CFT 
correspondence to the full string theory ({\em i.e.}, without
taking the scaling limits as in \cite{malda}), the closed string
degrees of freedom including gravity are no longer decoupled from
the effective theory on the world-volume of the branes.
To account for the closed string degrees of freedom, one introduces 
into the AdS supergravity
a hypothetical Planck brane with dynamical degrees 
of freedom representing these closed string modes. 
This hypothetical Planck brane is placed  
at the edge of the AdS throat created by the branes and effectively 
cuts off the radial AdS coordinate.
In the AdS/CFT correspondence, 
distances from the brane correspond to 
energy scales in the worldvolume theory.
Therefore, the Planck brane serves as an UV cutoff
and quantum gravity effects become important 
as we get closer to the Planck brane.

In the
 case that the spacetime transverse to the branes is compactified, 
the information about the compactification geometry
would then be encoded in vacuum expectation values 
of the excitations of the Planck brane. In principle one could derive 
this set-up by integrating out coordinate shells of constant warp factor 
(momentum shells on the world volume) extending beyond the throat of the 
AdS near horizon region.

\section{The Low Energy Spectrum}\label{moduli}
 
The low energy spectrum of
warped compactifications is naively different from that of
product space compactifications where one may use topological
information of the internal manifold to determine the massless
spectrum. For example the equation of motion of a scalar field,
\bq
\ove{\sqrt{g}}\pa_M \sqrt{g}g^{MN} \pa_N \phi(z) = 0 ~,
\fq
when reduced on a generic warped metric of the form 
\bq
g_{MN}(z) = \left(\matrix{ e^{2A(y)}\tilde{g}_{\mu\nu}(x) &0\cr
0& \bar{g}_{ab}(y)}\right)
\fq
yields 
\bq
(e^{-2A}\Box_{\tilde{g}} + \Box_{\bar{g}} +
d\bar{g}^{ab}\pa_aA\pa_b)\phi(z) =0 ~.
\fq
Here $d$ denotes the dimension of the internal manifold;
$D$ is the dimension of the ambient space.
Redefining $\phi(z) =e^{-dA/2}\tilde{\phi}(z)$ one finds that the low
energy modes descending from this field are determined by
\bq
\left(e^{-2A}\Box_{\tilde{g}} + e^{-\frac{dA}{2}}\left[\Box_{\bar{g}}
-\frac{d}{2}\Box_{\bar{g}}A-\frac{d^2}{4}(\pa A)^2\right] \right)\tilde{\phi}(z) =0 ~.
\label{ex1}
\fq
The last two terms act as additional mass terms for the dimensionally reduced
field. The massless modes are those in which 
the eigenvalue of the internal
Laplacian cancels the terms descending from the warp factor. 
Naively one thus loses the power of topological arguments to determine the
massless spectrum.
 
The particular warped solutions known in string theory/supergravity,
the
$p$-brane metrics, belong to a special class, however. The internal
manifold itself is also multiplied by a warp factor which precisely
compensates for the warp factor of the external space. For example, in
the scalar field above, the effect of an 
extra warp factor on the internal space, 
\bq 
g_{MN} = \left(\matrix{ e^{2A(y)}\tilde{g}_{\mu\nu}(x) &0\cr
0& e^{-2bA}\bar{g}_{ab}(y)}\right)~,
\fq
changes the field equation for the low-energy modes $\phi$ to
\bq
(e^{-2A}\Box_{\tilde{g}} + \Box_{\bar{g}}+(d-b(D-d-2))\bar{g}^{ab}\pa_aA\pa_b )\phi(z) =0 ~.
\label{ex2}
\fq
This internal warp factor
can cancel the additional mass terms in \rf{ex1} if $b$ is
chosen appropriately: $b = d/(D-d-2)$. This is in fact 
exactly the combination one
finds for $p$-brane metrics in supergravity. Here,
the physical reason is that the warp
factor is solely due to the backreaction of the vacuum
configuration. Indeed one can explicitly see from the solution that
the non-constant terms in the warp factor are suppressed by powers of
the gravitational coupling constant. In the limit where this vanishes
the warp factor is trivial and the space is of the product form $M_4
\times CY_4$. This is another argument why also in the case with
background fluxes, the integration constant
equals unity. In fact, the ``balancing'' of the warp factors is essentially
the reason why when we first
quantize open strings with Dirichlet-boundary conditions,
the gravitational backreaction of the D-branes does not change
the corresponding massless closed string spectrum.

The argument that in ``balanced'' warp metrics the warp factor is
to a large extent
inconsequential as regards to the massless part of the low-energy spectrum
holds irrespective of whether $G$-flux is present
or not. The introduction of the latter does complicate the
determination of the massless spectrum as in the presence of
non-trivial background expectation values of matter fields the mass
matrix of low energy modes is generically off-diagonal in terms of the
original fields. Fortunately the fact that the allowed background
fluxes are subject to topological conditions - $G$ should be primitive
and a (2,2) form - in addition to the topological interpretation of the moduli of the
Calabi-Yau four-fold ought to allow one to also use topological
methods to determine the massless spectrum with $G$-flux
\cite{Sethi}. In general these low-energy modes could be a
linear combination of the original fields, but their number can be
determined by looking at the topological constraints.   

To be specific, those Kahler moduli which spoil the primitivity condition
 $J\wedge G =0$ are lifted as well as those complex structure moduli
 which fail to keep $G$ a (2,2) class. In addition those Wilson lines
 which are not orthogonal to $G$: $C\wedge G \neq 0$ are lifted as
 well \cite{Sethi}. In principle one has hereby computed the
 massless spectrum. In practice, the first and the last constraint are
 readily solved but the counting of those complex structure
 deformations which keep $G$ a (2,2) class is more involved.
Consider a deformation of the complex structure. This is given by a
 coordinate transformation which is not holomorphic
\bq
z^i \rar y^i(z,\bar{z}) ~.
\fq
Infinitesimally the mixed and the pure deformations of the metric
 under an arbitrary coordinate transformation are 
\bqr
\del g_{a\bar{b}} &=&
 g_{a\bar{c}}\bar{\pa}_{\bar{b}}\bar{y}^{\bar{c}}+g_{\bar{b}c}\pa_a y^c + y^c \pa_c g_{a\bar{b}} +
 \bar{y}^{\bar{c}}\bar{\pa}_{\bar{c}} g_{a\bar{b}} ~, \non 
\del g_{ab} &=& g_{a\bar{c}}\pa_{b}\bar{y}^{\bar{c}}+g_{{b}\bar{c}}\pa_a
\bar{y}^{\bar{c}} \equiv g_{\bar{c}(a} \bar{\chi}^{\bar{c}}_{b)} ~. 
\fqr
This is just the well known fact that
non-holomorphic coordinate changes correspond to pure type
metric deformations. 
By contracting $g^{\bar{c}a}\del g_{ab} \equiv h^{\bar{c}}_b$ 
with the constant anti-holomorphic
 $(0,n)$ form one recovers the $(1,n-1)$ forms that are
 in one-to-one correspondence with complex structure deformations.
Under such a transformation a (2,2) form transforms infinitesimally as
\bq
G \equiv G_{a\bar{b}c\bar{d}} dz^ad\bar{z}^{\bar{b}}dz^cd\bar{z}^{\bar{d}}
~~~ \rar~~~ G+K^{(3,1)}+K^{(1,3)}
\fq
with
\bq
K^{(1,3)} = G_{a\bar{b}c\bar{d}}({\chi}^{a}_{\bar{e}}
 \del^{{c}}_{{e}}-
 {\chi}^{{c}}_{\bar{e}}\del^{{a}}_{{e}}) dz^ed\bar{z}^{\bar{b}}d\bar{z}^{\bar{d}}d\bar{z}^{\bar{e}}
~.
\label{htt}
\fq
For those complex structure 
deformations which keep $G$ a (2,2) class $K^{(1,3)}$ must
 vanish. This constraint can be expressed as follows. 
The complex structure deformations $\chi^a_{\bar{b}}$ are representatives of the cohomology $H^{(0,1)}(T)$. Define the form $\tilde{G} ~\eps~ H^{(0,2)}(\wedge^2{T})$
\bq
\tilde{G}_{\bar{c}\bar{d}}^{ab} = G_{\bar{c}\bar{d}ef} \Ome^{{e}{f}{a}{b}}~.
\label{lift}
\fq
The constraint that $K^{(3,1)}$ must vanish can then be 
 expressed by requiring that the triple intersection numbers 
\bq
\int \Ome_{abcd}
\left({\chi}_{(0,1)}^{[a} \wedge \tilde{G}_{(0,2)}^{bc} 
\wedge \tilde{\alp}_{I\,(0,1)}^{d]} \right) \wedge \Ome_{(4,0)} = 0
\label{er}
\fq
vanish for all basis elements $\tilde{\alp}_I$ of $H^{(0,1)}(T)$. The latter are constructed from basis elements $\alp_I$ of $H^{(3,1)}(CY_4)$ 
as in \rf{lift}. Equation \rf{er} thus represents that
the natural inner product of $K^{(1,3)}$ with an arbitrary element $\alp_I$ 
vanish.
There is however no general analysis of this condition
if the (2,2)
form $G$ is also required to be integral.
One
 therefore has to solve \rf{er} on a case by case basis. Below we will
 do so for the simple example where the $CY_4$= $K3 \times K3$. 

The deformations that do not respect the constraints
should give rise to mass terms when we perform a KK reduction. 
By our previous reasoning we can to first order
ignore the effects of the warp factor. Then, we are on a product space
and the KK reduction 
is straightforward, with the caveat that the field equations are
now not satisfied. One finds that the
kinetic term for $C$ is responsible for 
mass terms for the pure and mixed type
metric fluctuations (numerical coefficients
are suppressed)  
\bq
m^2 h^2  \sim \int
h^{ab}G_{a\bar{b}c\bar{d}}G_{b\bar{c}}^{~~~c\bar{d}}h^{\bar{b}\bar{c}}
+
h^{a\bar{b}}G_{ab\bar{e}\bar{d}}G_{\bar{b}\bar{c}}^{\bar{g}\bar{e}}
h^{b\bar{c}}  + G_{a\bar{b}c\bar{d}} h^{\bar{d}\bar{k}}h_{\bar{k}}^{{e}}G_{{e}}^{~~a\bar{b}c}+G_{a\bar{b}c\bar{d}} h^{\bar{d}k}h_{{k}}^{{e}}G_{{e}}^{~~a\bar{b}c} ~.
\label{mmss}
\fq
In the first and the third term one recognizes the ``square'' of $K^{(3,1)}$ 
in \rf{htt}. For massless deformations it must
vanish.
 
 One may compare this   
with the conjectures made in \cite{gukov}. There
it was argued that the complex structure deformations have a
superpotential
\bq
W_{C} = \int \Ome \wedge G ~,
\label{GukCom}
\fq
where $\Ome$ is the holomorphic (4,0) form. 
For the calculation of the mass terms involving the complex structure
deformations, the relevant parts of the first and second deformation
 of the four-form $\Ome$ are 
\bqr
\left( \del\Ome \right)_{abc\bar{n}} &\sim & \Ome_{abcd}g^{d\bar{m}}h_{\bar{m}\bar{n}} + \ldots~,\non
\left( \del^2 \Ome \right)_{ab\bar{n}\bar{f}} &\sim & \Ome_{abcd}g^{c\bar{m}}h_{\bar{m}\bar{n}}g^{d\bar{e}}h_{\bar{e}\bar{f}}+ \ldots
\fqr
The non-trivial quadratic fluctuations in the superpotential are thus given by
\bq
\del^2 W = \int \del^2\Ome  \wedge G \sim 
\int \sqrt{g} h_{\bar{m}}^nh_{\bar{c}}^d\Ome_{abdn}G^{ab\bar{c}\bar{m}}~,
\fq
where we have used the fact that the background value of $G^{(2,2)}$ is selfdual. Reducing to $N=0$ components and noting that the $W^2|_{\th=0}$ term does not contribute, we find the mass term for the pure deformations
\bq
m^2 h_{zz}^2 \sim \int h_{\bar{c}}^d\Ome_{abdn}G^{ab\bar{c}\bar{m}} G^{np\bar{q}\bar{r}}\bar{\Ome}_{\bar{r}\bar{q}\bar{s}\bar{m}}h^{\bar{s}}_{p} ~.
\fq
Using the relation
\bq
\Ome_{abcd}\bar{\Ome}_{\bar{a}\bar{b}\bar{c}\bar{d}} \sim \eps_{abcd\bar{a}\bar{b}\bar{c}\bar{d}} ~,
\fq
the mass term reduces to
\bq
m^2h_{zz}^2 \sim \int h_{\bar{s}}^dG_{\bar{q}\bar{r}nd} G^{np\bar{q}\bar{r}}h^{\bar{s}}_{p}+h_{\bar{q}}^dG_{\bar{s}\bar{r}nd} G^{np\bar{q}\bar{r}}h^{\bar{s}}_{p}~,
\fq
which is of the same form as the first and third
mass terms from the Lagrangian in \rf{mmss}.

For the Kahler deformations 
the authors of \cite{gukov} conjectured the superpotential 
\bq
W_{K}= \int {\cal K} \wedge {\cal K} \wedge G ~,
\fq
where $K$ is the complexified Kahler form ${\cal K}=J+iB$. The reduction to $N=0$ components in this case is more involved. This 
will be reported elsewhere.

\section{An Example: $K3 \times K3$}\label{tops}

 In this section, we will compute the spectrum for the simple
example where the $CY_4$ equals $K3 \times K3$.
We will first determine the number of complex and
Kahler moduli by explicitly 
constructing the complete parameter space of solutions 
to the $G$-flux constraints on the orbifold
limit of $K3 \times K3$.  This allows us to simply count the number of
moduli by hand. At the end we will compare it to the topological
determination of allowed deformations outlined above.

\subsubsection*{\it Vanishing Fluxes}\label{branespectrum}

Before proceeding let us briefly recall for comparison 
the spectrum of M theory on a
$K3 \times K3$ without 
G-flux. In this case there are $H^{(3,1)}+H^{(2,1)}$ $N=1,d=4$ chiral
multiplets, corresponding to deformations of the complex structure and
Wilson lines of the three-form respectively, and $H^{(1,1)}$ $N=1,d=4$
vector fields, corresponding to Kahler deformations. 
For $CY_4$ =  $K3 \times K3$ the Hodge diamond equals 
\bqr
\matrix{&&&& h^{0,0}&&&&\cr
&&&h^{1,0}&&h^{0,1}&&&\cr
&&h^{2,0}&&h^{1,1}&&h^{0,2}&&\cr
&h^{3,0}&&h^{2,1}&&h^{1,2}&&h^{0,3}&\cr
h^{4,0}&&h^{3,1}&&h^{2,2}&&h^{1,3}&&h^{0,4}}
=
\matrix{&&&& 1&&&&\cr
&&&0&&0 &&&\cr
&&2&&40&&2&&\cr
&0&&0&&0&&0&\cr
1&&40&&404&&40&&1}
\fqr
The $K3 \times K3$ compactifaction is non-minimal in that it 
preserves four more supercharges than
the minimal four required by $N=1,d=4$ supersymmetry. Indeed we see
that the spectrum is given by 40 $N=2,d=4$ vector multiplets.

Taking one of $K3$'s to be elliptically fibered we can lift this
M theory compactification to F theory on $K3 \times K3$, which
is related to a number of other compactifications by a chain of 
dualities \cite{SVW,sadov,KST}. Since F theory on $K3$ is heterotic on $T^2$,
compactifying both sides on another K3 gives heterotic on $K3 \times T^2$.
On the other hand, heterotic on $K3$ is F theory on a Calabi-Yau
threefold $CY_3$. As a result, F theory on $K3 \times K3$ is
dual to F theory on $CY_3 \times T^2$.  

The orientifold dual of the above F theory
compactification can be found via Sen's map \cite{sen}. We take
the orbifold limit $K3=T^4/ \ZZ_2$ for each $K3$.
Let $(z_1,z_2,z_3,z_4)$ be the
complex coordinates of $T^8$, with $z_4$ being the fiber coordinate.
The orientifold dual is given by Type IIB on 
$T^6/ \{ R_1 \times (\Omega (-1)^{F_L} R_2) \}$
where $\Omega$ is the worldsheet parity operation, $R_1$ and
$R_2$ act as follows:
\begin{eqnarray} 
R_1 z_{1,2} &=& - z_{1,2}~, \quad \quad R_1 z_3 = z_3 ~,\\
R_2 z_{1,2} &=& z_{1,2}~, \quad \quad R_2 z_3 = - z_3 ~. 
\end{eqnarray}
The resulting orientifold dual is therefore 
the Gimon-Polchinski model \cite{GP,sagnotti} further compactified on
a $T^2$ and then T-dualized in the $T^2$ directions, so that instead of 
$5$-branes and $9$-branes, we have
$3$-branes and $7$-branes.
At a special point of the moduli space where $\tau$ is constant,
{\em i.e.}, four $D7$-branes are placed on top on each $O7$-plane,
the gauge group from the $7$-branes is $U(4)^4$. In addition, there are two
${\cal N}=2$ 
hypermultiplets in the ${\bf 6}$ representation of each $U(4)$ from the
$77$ open strings. The $3$-branes give rise to additional gauge symmetries.
In the case of maximal gauge symmetries, {\em i.e.} the $3$-branes are
sitting on top of each other,
the gauge group from the $3$-branes is $U(16)$.
The $33$ open strings also give rise to
two ${\cal N}=2$
hypermultiplets in the ${\bf 120}$ representations
of $U(16)$. Finally, there is a hypermultiplet in the bi-fundamental 
representation $({\bf 16},{\bf 4})$ of $U(16) \times U(4)$ if
the $3$-branes sit on one of the four groups of $7$-branes.
The orientifold model at a generic point of the moduli space
can be obtained from the above by Higgsing.

\subsection{G-flux conditions}\label{quantization}

Now we turn on background fluxes. The field-equations demand that we
 seek an integer (2,2) form on $K3 \times K3$ = $T^4/\ZZ_2
\times T^4/\ZZ_2$ which
is primitive, i.e. 
\bq
J \wedge G =0~. 
\label{prim}
\fq
We will choose $G$ of
the form  
\bq
G = \ome_1 \wedge \ome_2~,
\fq
where $\ome_i ~\eps ~H^{(1,1)}(T^4/\ZZ_2) \cap H^{(2)} (T^4/\ZZ_2,\ZZ)$. 
This guarantees that $G\, \eps
\, H^{(2,2)}(T^4/\ZZ_2,\ZZ)$ though it is not necessary. 
This simplifies the solutions to the
quantization conditions (as compared to the general case \cite{Sethi}).
We could have chosen
$\ome_i ~\eps ~H^{(2,0)}(T^4/\ZZ_2,\ZZ)$ or
$H^{(0,2)}(T^4/\ZZ_2,\ZZ)$ but these forms correspond to the class of the
fiber and its Hodge dual and are no longer normalizable 
when we shrink the volume of the fiber in the 
F theory limit \cite{Sethi}. 

The complex structure is inherited from the tori and the condition
\rf{prim} requires that each $\ome_i$ is primitive with respect to
$J_i$ \cite{Sethi}. 
Thus our task is reduced to finding primitive (1,1) forms on
$T^4/\ZZ_2$. The cohomology $H^{(1,1)}(K3=T^4/\ZZ_2)$ has dimension 20, but
these can be subdivided in four ``untwisted'' (1,1) forms inherited
from the $T^4$ and sixteen ``twisted'' ones. As any integer form is an
integer combination of the basis forms we can consider the two
situations separately.  

\subsubsection*{\it Constant fluxes}

Consider untwisted forms first. These are inherited from the $T^4$
and are constant on the orbifold. The intersection matrix of
two-forms on K3$=T^4/\ZZ_2$ is block-diagonal with in the upper
left-hand corner minus the Cartan-matrix of $E_8 \times E_8$ and in
the lower right-hand corner three times the Pauli matrix $\sig_1$. The
latter blocks equal the intersection-matrix of two-forms on $T^4$ and
thus correspond to the untwisted forms. Hence we only have to check that
the periods of the untwisted forms are integer over the untwisted
cycles. This means that we have reduced our problem to finding the set of 
integer (1,1) forms $D$ on $T^4$.    
   
For each $T^2$ (with volume normalized to 1),
we make the standard choice for the periods 
\bq
\int_{\gam_x^j} dz^i = \del^i_j \hspace{0.7in} \int_{\gam_y^j} dz^i =
\tau_i\del^i_j \hspace{0.7in} \mbox{(no sum on $i$)}~,
\label{whopper}
\fq
where $\gam_x^i$ and $\gam_y^i$ are the $x$ and $y$ cycles; $i,j=1,2$. The 
(1,1) forms $\tilde{\gam}_x^i$ and $\tilde{\gam}_y^i$ dual to these
cycles are
\bq
\tilde{\gam}_x^i = dy^i \hspace{1in} \tilde{\gam}_y^i =
-dx^i ~,
\fq
with the volume element
\bq
\int_{T_i} dx^i\wedge dy^i = 1 \hspace{0.7in} \mbox{(no sum on $i$)}~.
\fq
The Kahler form is given by
\bq
J_i = dz^i \wedge d\bz^i~,
\fq
and equals $(\bar{\tau}_i - \tau_i)$ times the volume form $dx^i\wedge
dy^i$. 

Take the Kahler form $J$ for 
the $T^4$ to be the sum of $J_1$ and $J_2$. The
general form of (1,1) forms obeying the primitivity condition is then
\bq
J\wedge D = 0  \rar D= Adz_1d\bz_2+\bar{A}d\bz_1dz_2 
+i B(dz_1d\bar{z}_1-dz_2d\bar{z}_2)~.
\label{ert2}
\fq
where $A$ is complex and $B$ is real.
The requirement that $D ~\eps ~H^{(1,1)}(T^4,\ZZ) \subset
H^{(1,1)}(T^4/\ZZ_2,\ZZ)$ 
demands that
\bqr
A+\bar{A} &=&n ~,\non
\bar{\tau}_2A+\tau_2 \bar{A} &=& m ~,\\
\tau_1A +\bar{\tau}_1\bar{A} &=& p ~,\non
\tau_1\bar{\tau}_2 A + \bar{\tau}_1\tau_2 \bar{A} &=& q~, \nonumber
\fqr
where $n,m,p,q ~\eps~ \ZZ$. In addition
\bqr
i B(\bar{\tau}_1-\tau_1) &=& v ~,\non
i B(\bar{\tau}_2-\tau_2) &=& w ~,
\label{bbbb}\fqr
where $v,w ~\eps~ \ZZ$. These are six equations with in principle
three unknowns: Re$(A)$, Im$(A)$, $B$; the system is overconstrained and
has no solutions. 

However, if one relaxes the requirement that all the $\tau_i$
are free parameters, namely one allows the complex structure $\tau_2$ of
one torus to be determined in terms of the other, and in 
addition requires that the ratio of the 
imaginary parts of $\tau_1$ and $\tau_2$ be a rational number, then there
is a solution.
This poses three real constraints on the moduli of the
$T^4$. 
The first constraint can be seen by rewriting the integrality conditions as
\bqr 
\bar{A} &=& n-A ~,\non  A &=&
\frac{p-n\bar{\tau}_1}{(\tau_1-\bar{\tau}_1)} =
\frac{m-n{\tau}_2}{(\bar{\tau}_2-{\tau}_2)}  ~,\label{quant}
\\ 
\frac{\tau_1-\bar{\tau}_1}{\tau_2-\bar{\tau}_2} &=&
\frac{n\bar{\tau}_1-p}{m-n{\tau}_2} ~,\non
\tau_1\bar{\tau}_2 A + \bar{\tau}_1\tau_2 \bar{A} &=& q~. \nonumber
\fqr
Note, however, that we have imposed both the primitivity
and integrality conditions in deriving these constraints. Moreover, 
in (\ref{ert2}) we are only considering (1,1) classes. 
Hence, the three constraints determine the loci in the combined 
space of complex and Kahler structures 
on K3 where one may find primitive integral forms which are purely (1,1). 
The number of constraints therefore correspond to the total number of moduli, 
complex structure plus
Kahler, which are lifted. Complex structure deformations must form complex 
pairs. The total number of moduli which become massive is therefore 1 
(complex valued) complex structure deformation and 1 real Kahler deformation.
For instance, one can see this explicitly by generalizing the choice of
the Kahler class to $J_1+aJ_2$ and then noting that the analogue
of eq.(\ref{bbbb}) fixes the value of $a$ in terms of the complex
structures $\tau_1$ and $\tau_2$.

In the end we are interested in the $H^{(2,2)}(T^4,\ZZ)$ flux 
\bq
\int_{T^4} D\wedge D = 2(|A|^2-B^2)(\tau_1-\bar{\tau}_1)(\tau_2-\bar{\tau}_2)~.
\label{fll}
\fq
which should be an integer. 
This is proportional to the number of M2/D3-branes we will
have to introduce. Substituting the second equation of \rf{quant} in
\rf{fll} we
find that
\bq
\int_{T^4} D\wedge D = -2vw-2(p-n\bar{\tau_1})(m-n\bar{\tau_2})~.
\fq
As the last factor should be a negative semi-definite integer ($\tau_i
~\eps~ H^+$ in \rf{fll}) let us make the Ansatz that
\bqr\label{tau1tau2}
m-n\bar{\tau_2} = \frac{r}{(p-n\bar{\tau_1})} &\rar& \bar{\tau}_2 =
\left(\frac{m}{n} -\frac{r}{n(p-n\bar{\tau_1})}\right)
\fqr
is indeed a solution to \rf{quant}. Here $r~\eps~ \ZZ^+/2$. Substituting
this relation into the four equations \rf{quant} we find that
our Ansatz is a solution with $r=-qn+mp$. Note that $mp \geq
qn$. Hence the flux equals
\bq
\int_{T^4} D\wedge D = -2vw-2(p-n\bar{\tau_1})(m-n\bar{\tau_2})
= 2(qn-mp-vw)~.
\fq
Note that if  $D$ is also an element of $H^{(1,0)}(T^2,\ZZ)\times
H^{(0,1)}(T^2,\ZZ) \subset H^{(1,1)}(T^4,\ZZ)$, i.e. if $n=n_1n_2$; $m=n_1m_2$; $p=m_1n_2$ and $q=m_1m_2$ for
integer $n_1,n_2,m_1,m_2$ then the flux $\int_{T^4}D\wedge D
=2(m_1m_2n_1n_2-n_1m_2m_1n_2)$ vanishes.

\subsubsection*{\it Twisted fluxes}

We will restrict ourselves to twisted forms whose dual cycles are
localized at the fixed points. These are two-speres $S^2 \simeq CP_1$
shrunk to a point. Denoting them as $B^i$ with $i=1,...,16$ running
over the fixed points, the general form of a twisted flux $D$ is
\bq
D = c_iB^i~.
\label{twist}
\fq
The $B^i$ are a linear combination of the generating forms $V^n$ of 
$H^{(1,1)}(T^4/\ZZ_2,\ZZ)$: $B^i = a^{i}_{n}V^n$ with $a_n^i$
integer. As we are limiting
our attention to 
twisted fluxes we may restrict the $V^n$ to those
generating minus the $E_8 \times E_8$ Cartan matrix $I^{nm}$. This also
means that 
$D$ is automatically primitive as the Kahler form $J$ consists purely
of untwisted forms. 

The condition that also $D~\eps~ H^{(1,1)}(T^4/\ZZ_2,\ZZ)$ means that
\bq
\int_{V^n} D = -c_ia^i_mI^{mn} =  p^n ~,
\label{quant4}
\fq
with $p^n~\eps~ \ZZ$ for all $n$. The cycles $B^i$ have an intersection 
matrix \cite{Strom}
\bq
\int B^i \wedge B^j = -2\del^{ij}=-a^i_nI^{nm}a^j_m~.
\fq
This means that the
coefficients $c_i$ are all multiples of $1/2$, 
\bq
c^i = -\frac{p^na^i_n}{2} ~.
\fq
The four-form flux of interest to us $\int_{T^4/\ZZ_2} D \wedge D$ equals
\bq
\int_{T^4/\ZZ_2} D \wedge D = -2\sum_i c^ic^i~ = - p^nI^{-1}_{nm}p^m~,
\label{tot}
\fq
as one can show that
\bq
\sum_i a^i_na^i_m = 2I^{-1}_{nm}~.
\fq
The inverse of the Cartan matrix has integer
entries because the $E_8$ lattice is even and self-dual, and
the diagonal entries are even.  
The r.h.s. of \rf{tot} is therefore an integer as
expected. Note again that $\int D\wedge D$ is negative semidefinite
just as in the case of constant fluxes.

Note that the $B_i$ or more precisely arbitrary half-integer
combinations thereof do not generate $H^{(1,1)}(T^4/\ZZ_2,\ZZ)$  
\cite{Strom}, but a larger group. For integer cohomology the $c^i$ are
subject to the additional constraint
\bq
\sum_i a^i_m c^i = p^n I\inv_{nm} ~.
\fq
We are therefore not missing any integer forms by limiting our
attention to fluxes of the form \rf{twist}.

In this case those Kahler deformations $\del J = \beta_nV^n$ where
$V^n~\eps~H^{(1,1)}(T^4/\ZZ_2,\ZZ)$ are twisted fluxes that do not
preserve the primitivity condition are frozen. Those deformations such 
that
\bq
\int \del J \wedge D =- \beta_n I^{nm}c_ia^{i}_{m} =0~~~~\forall~\beta_n
\label{ert}
\fq
do survive. The integrated primitivity condition is sufficient to
guarantee the local one \cite{Strom}.   

In fact we can determine the contribution to $\int D\wedge D$ of
each fixed point separately. As the intersection matrix of the cycles
$B^i$ is proportional to the unit matrix, the contribution of the
`$i$'th cycle is just given by the `$i$'th term in equation \rf{tot}
\cite{Strom}. Since the orientifold planes are also localized at the
fixed points twisted fluxes change the effective O-plane charge.

\subsection{The Spectrum}

In the previous section, we 
explicitly constructed the parameter space 
of solutions for pure $(1,1)$ forms on $K3$ that are primitive and
integral.
For constant flux this showed that the dimension of the space of complex 
structures and the dimension of the space of Kahler forms are each reduced 
by one. In this section we will compare these results
with a topological deformation analysis.
 
Suppose one is given a complex structure and Kahler form on $K3$ 
for which it is possible to find a primitive 
integral (1,1) flux $\ome_i$. The
$\ome_i \in H^{(1,1)} (K3) \cap H^2 (K3,\ZZ)$  define
a single direction in $H^{(1,1)} (K3) \cap H^2 (K3,\ZZ)$. 
The Kahler deformations which are preserved are those which are orthogonal
to $\ome_i$ in the sense that
\begin{equation}
\int \delta J \wedge \ome_i =0 ~.
\end{equation}
Thus there is always
exactly 
one Kahler deformation in each K3 that is lifted.

As for the complex structure deformations, on a $K3$ they will cause
a (1,1) form to become a mixture of  
a $(1,1)$ and a $(0,2)$+$(2,0)$ form; the latter being 
complex conjugates. 
Since $H^{(0,2)}$ is 1 dimensional, a 19 dimensional subspace of
(1,1) forms is preserved. These allowed
deformations will not spoil integrality, but could spoil primitivity. One can, however, always find a compensating Kahler deformation which restores 
primitivity. This is illustrated in the constant flux example by our
earlier explicit calculation.

Since on $K3 \times K3$, the (2,2) classes
come from $(1,1) \times (1,1)$ and $(2,0) \times (0,2)$,
we find that there is a
$38= 19 + 19$ dimensional subspace of preserved
complex structure deformations on $K3 \times K3$. 

An alternative way to see this embodies the symmetry between complex structure deformations and Kahler deformations. Consider in analogy with the superpotential \rf{GukCom} of \cite{gukov} the intersection 
\bq 
\int \Ome_i \wedge \ome_i
\fq
for each $K3$. $\Ome_i$ is now the holomorphic (2,0) form 
Under a complex structure deformation this potential changes to 
\bq
\int \del \Ome_i \wedge \ome_i ~.
\fq
We see that that deformation which is parallel to the $\ome_i$ flux is lifted.

These constraints thus change
the number of chiral and vector multiplets by one in each K3.
As a result, a total of two $N=2$,
$d=4$ vector multiplets are lifted due to the G-flux. 
The spectrum (in addition to that from the branes) thus consists of 
38 $N=2$, $d=4$ $U(1)$ vector multiplets, coupled to gravity.

\section{Discussion}\label{discussion}

In this paper, 
we have taken some modest steps towards understanding
warped compactifications in M and F theory on Calabi-Yau
four-folds in the presence of non-trivial background flux. The 
introduction of background flux can have interesting
consequences for low energy physics and 
this subject is certainly
worthy of further study.
Detailed investigations require
the explicit form of the warp factor, and
to facilitate its determination, 
we considered
orbifold limits of these compactifications.
The contribution of the background flux is a
backreaction effect similar to explicit $p$-brane sources and because the
energy density associated with the G-flux is inversely proportional to 
the volume, the leading term
in the warp factor is always a constant. This contrasts sharply to what
has
been found in a number of other warped scenarios.
In the orbifold limit
the background fluxes are either constant or localized. The
constant fluxes allow the warp factor to be consistently
solved in terms of Green's functions on the internal space.
The twisted
fluxes
act as sources for the Green's functions.

The extensive nature of the background flux suggests 
that its introduction will not have drastic consequences;
indeed we find 
that the usual relation between the four- and ten-dimensional Planck 
scales is recovered.
This, again
contrasts with
the scenarios involving a cosmological constant \cite{RS}, 
in which gravity is localized.
The shape of the graviton wavefunction is sharply peaked 
at the location of the branes but this is just the source 
divergence of a Green's function. 
In a similar analysis for
Heterotic-M theory on a Calabi-Yau
three-fold, discussed  in an appendix,  we also derived
the wavefunctions of the KK modes
In our M/F theory setup,
the higher KK modes can also
be obtained
by linearizing gravity if the compactification size is large. 
Interestingly, aside from the compactification size,
at subleading order the masses
of the KK modes appear to depend on the location of the branes.

In the view of \cite{hv}, in which the renormalization
flow of the effective world volume theory is correlated with
a translation in one of the real internal directions,
the compactification geometry is encoded by a hypothetical
Planck brane at the throat of the near-horizon AdS created by the
branes.
The shape of the warp factor should determine the characteristics
of this Planck brane.

Finally we discussed the low energy spectrum of
these warped compactifications. Supersymmetry or the argument that the
warp factor is solely due to backreaction guarantees that topological
arguments may still be used to determine the massless spectrum.
The presence of background fluxes can lift a number of moduli, but the
rules to determine them are of (pseudo) topological nature as well. For
$K3 \times K3$
we found that  two $N=2$ vector multiplets were lifted.
In addition to the complex and Kahler moduli that
arise in these compactifications, the low energy degrees of freedom
include gauge and matter fields on the branes. For example,
in the $K3 \times K3$ case,
there are generically $3$ and $7$ branes on which gauge and matter
fields
are localized. We recalled in Section \ref{branespectrum} the
spectrum from the branes when the background flux is zero.
The determination of the
spectrum from the brane sector of F theory compactifications
in the presence of
background three-form NS-NS and R-R fluxes
is an interesting direction to pursue in the future.

\medskip

\noindent {\bf Acknowledgments}

\smallskip

We would like to thank Jan de Boer, Claude LeBrun, Lennaert Huiszoon, Zurab Ka\-ku\-shad\-ze, David Morrison,
Horatiu Nastase,
Peter van Nieuwenhuizen and Savdeep Sethi
for discussions and correspondence.
The research of G.S. is partially supported by the NSF grant PHY-97-22101,
and that of B.R.G. is partially supported by the DOE grant DE-FG02-92ER40699B.
K.S. acknowledges the hospitality of the C.N. Yang ITP and the Spinoza Institute.

\section*{Appendix}
\subsection*{A. Green's Functions on a compact space}

On a compact space, or equivalently when the Laplacian has
non-trivial zero modes, the inversion of the Laplacian
is only defined on the subset of
functions orthogonal to these zero-modes.
One can see this explicitly by constructing the Green's function in
terms of the complete set of eigenfunctions of the
Laplacian. These eigenfunctions must obey the same boundary
conditions as the function on which the Laplacian acts. Recalling that the
Laplacian has non-positive definite eigenvalues we can denote an
orthonormal set by $\phi_m(x)$ with 
\bq 
\Box \phi_m(x) = -m^2\phi_m(x) ~.
\fq 
The Green's function is then 
\bq 
G(x,x\pr) = -\sum_{m\neq 0}
\frac{\bar{\phi}_m(x) \phi_m(x')}{ m^2} ~.
\fq 
All the
zero modes of the Laplacian must be omitted in the sum on the
r.h.s. This is a generic feature of all Green's functions. When one writes 
\bq 
\Box G(x,x\pr) = \one =\mbox{``}\del^D(x-x\pr)\mbox{''} ~,
\fq 
it is implicitly understood that the
$\delta$-function or identity-operator is only defined in the space of
functions orthogonal to the zero-modes of the Laplacian. This is not
necessarily the same as the Dirac-delta function. On a circle of radius $R$
for instance, there is exactly one non-trivial zero-mode, the constant
function. The Green's function equals
\bq 
G(x,x\pr) = - \ove{2\pi R}
\sum_{n\neq 0} \frac{e^{\frac{in(x-x\pr)}{R}}}{n^2/R^2} 
\fq 
and therefore obeys\footnote{Recall that we use the convention that
the Dirac-delta function in curved space obeys $\int dx \sqrt{g(x)} f(x)
\delta (x-x_i) = f(x_i)$.} 
\bq 
\Box G(x,x\pr) = \pa_x^2 G(x,x\pr) = \ove{2\pi R}\sum_{n\neq
0} e^{\frac{in(x-x\pr)}{R}} = \del(x-x\pr) - \ove{2\pi R} ~,
\label{1dim}
\fq 
rather
than 
\bq 
\Box G(x,x\pr) = \del(x-x\pr) ~.
\fq 
The former combination
$\del(x-x\pr) - \ove{2\pi R}$ is obviously the delta-function in the
space of functions orthogonal to the zero-mode. In the infinite volume
limit we are left with the conventional Dirac-delta function.

The above is a reflection of the fact that for differential forms, 
the Green's function can be used as a projector onto the space of
non-harmonic forms, see e.g. \cite{Eguchi}.  An
arbitrary $p$-form $\ome$ can be decomposed into an harmonic $p$-form
$\gam$ and a non-harmonic part as 
\bq
\ome = \gam +\Box G \ome~.
\fq
The combination $\Box G$ is equivalent to the projector 
\bq
\Box G = \one - \sum_n \frac{\gam_n} {\left(\int \ast
\gam_n \wedge \gam_n\right)} \int \ast
\gam_n \wedge~.
\fq 
Here the $\gam_n$ form a 
basis of harmonic $p$-forms. For
scalars on a compact manifold there is only one harmonic form, the
constant, and for 0-forms the above combination reduces to 
\bq 
\Box G(x,x\pr) = \del^D(x-x\pr) - \ove{\mbox{Vol}} ~.
\fq
For one dimension this equals the r.h.s. of   
\rf{1dim}

Particularly in the case of tori, one occasionally imposes the boundary
conditions signifying the compactness of the space 
by introducing image charges on the covering
space. The torus is considered as the quotient $\mathbb{R}^d/ \Lam_d$
of $\mathbb{R}^d$ by the lattice $\Lam_d$. The Green's function on
$\mathbb{R}^d/ \Lam_d$ is then a sum of regular $\mathbb{R}^d$ Green's
functions obeying 
\bq
\Box G_{\mathbb{R}^d}(x,x') = \del_{Dirac}^D(x-x\pr)
\fq
such that $G_{\mathbb{R}^d/\Lam_d}$ and its derivative is periodic, i.e.
\bq
G_{\mathbb{R}^d/\Lam_d}(x,x') = \sum_{\vec{n}}
G_{\mathbb{R}^d}(x,x\pr+\vec{n}\cdot \vec{e})~.
\fq
Here $\vec{e}$ are the lattice vectors generating $\Lam_d$.

The Laplacian acting on $G_{\mathbb{R}^d/\Lam_d}(x,x')$ now gives
\bq
\Box G_{\mathbb{R}^d/\Lam_d}(x,x') =
\sum_{\vec{n}}\del^D_{Dirac}(x-x\pr+\vec{n}\cdot \vec{e})~.
\fq
The only relevant part of the r.h.s, however, is the term with $n=0$ as both
$x,x\pr$ belong to a single fundamental region $x^i\,\eps\,[0,e^i)$. The delta functions for other values of $\vec{n}$
never contribute. 

If one chooses to solve for the warp factor using such a Green's
function, one would {\em ignore} the
contribution from any constant terms on the r.h.s. 
This is evident from the preceding discussion. However, a
consistency condition is required, namely, the total
integral on the r.h.s. of the warp factor equation of motion must
vanish. 
This just says that the total flux on the compact space is zero
or that the anomaly cancels.

\subsection*{B. Heterotic-M theory on $CY_3$}

The effective five-dimensional theory by
compactifying Heterotic-M theory on a Calabi-Yau 3-fold was
derived in \cite{ovrut}. Here, we will follow their discussion and notation. 
We will consider
the simplest case, in which there are no 5-branes,
and the 
low energy degrees of freedom include only gravity and the hypermultiplet
containing the breathing
mode $V$ of the Calabi-Yau (the ``universal'' solution in \cite{ovrut}). 
With the standard embedding of the spin
connection in one of the $E_8$'s, the effective action 
becomes \cite{ovrut}
\bqr
 2\kap_5^2 S_5 &=& - \int_{M_5}\sqrt{-g}\left[R
+\frac{1}{2}V^{-2} g^{55} V'^2
         +\frac{1}{3}V^{-2} \alp^2
         \right] \nonumber\\
      && \qquad\qquad
         + 2\sqrt{2}\int_{M_4^{(1)}}\sqrt{-g}\, V^{-1}\alp - 2\sqrt{2}
         \int_{M_4^{(2)}}\sqrt{-g}\,V^{-1}\alp \; ~,
\fqr
where $\kappa_5$ is the five-dimensional gravitational coupling.
The constant $\alpha$ is given by
\begin{equation}
\alpha = - {1 \over {8 \sqrt{2} \pi v}} 
\left( {\kappa \over{4 \pi}} \right)^{2/3}
\int_{CY_3} \omega \wedge \mbox{tr}  R \wedge R ~,
\quad \quad 
v=\int_{CY_3} 
\sqrt{g_{CY_3}}~,
\end{equation}
where $\omega$ is the Kahler form on the $CY_3$.
The metric is warped to
\begin{equation}
ds^2 = a^2(y) \eta_{\mu \nu} dx^{\mu} dx^{\nu} + b^2(y) dy^2~.
\end{equation}
The solution to the field equations is:
\bqr
 a &=& a_0H^{1/2}~, \nonumber \\
 b &=& b_0H^2~,\qquad\qquad H=\frac{\sqrt{2}}{3}\alp |y|+c_0 \equiv 
c |y|+c_0\;~, 
 \label{u_sol}\\
 V &=& b_0H^3\nonumber \; ~,
\fqr
where $a_0$, $b_0$ and $c_0$ are integration constants.

The Einstein equation linearized in the fluctuations give: 
\begin{equation}
-{1\over 2} h''_{\mu \nu} 
- {1\over 2} \Big( 4~ {a^{\prime} \over a} - {b^{\prime} \over b} \Big) 
h'_{\mu \nu}
- \Big( 3c b_0 H -3 {c \over H} \Big) \Big( \delta(y) - \delta(y-\pi \rho) \Big) h_{\mu \nu}
= {1\over 2} ({b_0 \over a_0})^2 H^3 h_{\mu \nu, \lambda \lambda}~.
\end{equation}
Let us first consider the equation in the bulk.
It is easy to see that the above equation can be written as
the covariant 
wave equation (\ref{covariantwave}).
As in Section \ref{shape}, we expand 
$h_{\mu \nu} (x,y) = \hat{h}_{\mu \nu} (x) \psi (y)$ with
$\Box_{x} \hat{h}_{\mu \nu} (x) = m^2 \hat{h}_{\mu \nu} (x)$. 
The properly normalized wavefunction is
\begin{equation}
\Psi (y) = [e^{-2 A(y)} \sqrt{g(y)}]^{1/2} \psi(y)
=a b^{1/2} ~\psi(y)~.
\end{equation}
For massless graviton, 
\begin{equation}  
\Psi (y) \sim a b^{1/2} \sim H^{3/2}~.
\end{equation}
Therefore, the wavefunction $\Psi (y)$ vanishes at the
singularity where $H(y)$ is zero.

The
compactification scale of the M theory direction is usually taken
to be slightly
larger than that of the Calabi-Yau (from gauge and gravitational
unification \cite{horava}). 
Therefore, there is a regime in which the theory is five-dimensional, and
the wavefunction of some of the low-lying massive KK gravitons
can still be described by the above wave equation
\begin{equation}
-{1\over 2} \psi^{''} (y) 
- \Big( 3c b_0 H -3 {c \over H} \Big) \Big( \delta(y) - \delta(y-\pi \rho) 
\Big) \psi (y)
= {1\over 2} m^2 ({b_0 \over a_0})^2 H^3 \psi (y)~.
\end{equation}
 
Let us rewrite the above equation such that $m^2$ becomes the
eigenvalue of a Schrodinger equation. Define a new 
variable $u=(2/5)~c^{-1}~(b_0/a_0)~H^{5/2}$ and hence
$d u = (b_0/a_0) H^{3/2} d y$. In
terms of this
variable, the linearized equation becomes
\begin{equation}
-{1\over 2} \ddot{\psi} 
- {3 c\over 4} ({a_0 \over b_0}) H^{-5/2} \dot{\psi}
-3 a_0 c \left( {1\over H^{1/2}} - {1\over b_0 H^{5/2}} \right)
\Big(\delta(u-u_1) - \delta(u-u_2) \Big) \psi 
={1\over 2} m^2 \psi~,
\end{equation}
where the dot denotes the derivative
with respect to $u$. The locations $u_1$ and $u_2$ are
defined by $u=u_i$ when $y=0$ and $\pi \rho$ respectively.

Finally, to eliminate the first derivative term in the above
equation. Define $\psi = H^s \hat{\psi}$.
It is easy to see that first order derivatives of $\hat{\psi}$ do 
not appear if $s=-3/4$. The function $\hat{\psi}$ satisfies 
the following equation
\begin{equation}\label{schrodinger}
-{1\over 2} {{d^2} \over du^2} \hat{\psi} (u) + {\cal V} (u) \hat{\psi} (u) 
= {1\over 2} m^2 \hat{\psi} (u)~,
\end{equation} 
where
\begin{eqnarray}
{\cal V} (u) &=& -{21 \over 32} \left({a_0 \over b_0}\right)^2 {c^2 \over H^5} 
- \left( {a_0 \over b_0} \right)
{c \over H^{5/2}} \left( 3 b_0 H^2 - {15\over 4} \right)
\Big( \delta(u-u_1) -\delta(u-u_2) \Big) \nonumber  \\
&=& -{21 \over {200 u^2}}  -{2 \over {5 u}}\cdot
\left[  3 b_0 \left({5c \over 2} {a_0 \over b_0} u \right)^{4/5} -{15 \over 4}
\right]~\Big( \delta(u-u_1) -\delta(u-u_2) \Big) \nonumber~. 
\end{eqnarray}
The potential is not of the ``volcano'' type \cite{RS}.
In particular, we have seen that gravity is not localized.

We have seen that the wavefunction of the massless graviton
can be deduced without having to solve (\ref{schrodinger}) since
it is clear from (\ref{wave}) that $\psi=$  constant is
a solution. The massive modes can be found by
solving (\ref{wave}), or equivalently (\ref{schrodinger}).
Let us focus on the latter, as it allows us to compare our
solution with that of \cite{RS}. 
Let $\hat{\psi}=u^{1/2} f$,
the Schrodinger equation can be written as
\begin{equation}
v^2 {{\partial^2 f} \over {\partial v^2}} +
v {{\partial f} \over {\partial v}} +
(v^2 - {1 \over 25}) f (v)= 0~,
\end{equation}
where $v=m(u+u_0)$. The solution is simply
\begin{equation}
f= k_1 J_{1/5} \left( m(u+u_0) \right) 
+ k_2 J_{-1/5} \left( m(u+u_0) \right)~,
\end{equation}
where $k_1$, $k_2$ are constants which can be determined by
the matching conditions at the delta function sources.
The boundary conditions at the two delta sources also imply
that the masses $m$ of the KK modes are quantized in units
of $1/\rho$.

\end{document}